\documentstyle[12pt,epsfig]{article}
\newlength{\dinwidth}
\newlength{\dinmargin}
\def\docnum#1{\hbox to \hsize{\hskip123mm\hbox{#1}\hss}}
\def\date#1{\edef\@temp{#1}\ifx\@temp\@empty\def\@temp{\today}\fi
\hbox to \hsize{\hskip123mm\hbox{\@temp}\hss}}
\def\title#1{\vskip 0.8in plus 2in\begin{center}%
{\Large\bf#1\par}\vskip1.5em\end{center}\vskip 1in}
\def\@makefnmark{\hbox{$^{\@thefnmark)}$}}
\def\author#1{
\setcounter{footnote}{0}\def\@currentlabel{}%
\begingroup\def\thefootnote{\arabic{footnote}}
\def\@makefnmark{\hbox{$^{\@thefnmark)}$}}
\global\@topnum\z@ \large\begin{center}{\lineskip.5em
\begin{tabular}[t]{c}#1\end{tabular}\par}
\end{center}\par\vskip1.5em\@thanks\endgroup}

\def\abstract{\vskip0.8in plus 3in\begin{center}{\large\bf Abstract}\end{center}\quotation}

\begin{document}

\begin{titlepage}
\title{Hadron Production \\
 in\\
 Au-Au collisions at RHIC
 }

\vspace{-2.6cm}
\begin{center}
\large{P. Braun-Munzinger$^a$, D. Magestro$^a$, K.
Redlich$^{a,b}$,
and J. Stachel$^{c}$ }\\
\vskip 1.0 cm

{\small\it $^a$ Gesellschaft f\"ur Schwerionenforchung, D-64291
Darmstadt, Germany}
\\
{\small\it $^b$ Institute of Theoretical Physics, University of
Wroclaw,
PL-50204 Wroclaw, Poland}\\
{\small\it $^c$ Physikalisches Institut der Universi\"at
Heidelberg, D-69120 Heidelberg, Germany}\\
\end{center}

\begin{abstract}

We present an analysis of particle production yields measured in central Au-Au
collisions at RHIC in the framework of the statistical thermal model. We
demonstrate  that  the model extrapolated from previous analyses at SPS and
AGS energy  is in good agreement with the available experimental data at $\sqrt
s=130$ GeV implying a high degree of chemical equilibration. Performing a
$\chi^2$ fit to the data, the range of thermal parameters at chemical
freezeout is determined. At present, the best agreement of the model and the
data is obtained with the baryon  chemical potential $\mu_B\simeq 46\pm 5$ MeV
and temperature $T\simeq 174\pm 7 $
 MeV. 
 More ratios, such as  multistrange baryon  to meson, would be
required to further constrain
 the chemical freezeout conditions.
Extrapolating thermal parameters to higher energy, the predictions
of the model for particle production in Au-Au reactions at $\sqrt
s=200$ GeV are also given.
\end{abstract}
\end{titlepage}

\newpage

\section{Introduction}

The ultimate goal of ultrarelativistic nucleus-nucleus collisions
is to study the properties of strongly interacting matter  under
extreme conditions of high energy density \cite{satz,stachel1,stoc}.
 Hadron multiplicities  and their
correlations are  observables which can provide  information on
the nature,  composition, and  size of the medium from which they
are originating. Of particular interest  is the extent to which
the measured particle yields are showing equilibration. The level
of equilibrium of secondaries in heavy ion collisions can be
tested by analyzing the particle abundances
\cite{satz1,pbm1,pbm2,stachel2,cleymans1,prl,becattini}
or their momentum spectra \cite{pbm1,heinz2}. In the first case
one establishes the chemical composition of the system, while in
the second  case  additional information on dynamical evolution
and collective flow  can be extracted.

 A detailed analysis of
the experimental data at  SPS energy has shown that hadronic
yields and their ratios resemble those of a population in chemical
equilibrium. Most  particle multiplicities measured in
nucleus-nucleus collisions at the SPS are well consistent with
thermal model predictions \cite{pbm2,becattini,mark}. Particle
transverse mass distributions, on the other hand, exhibit slopes
showing a structures  typical  of a thermal source undergoing
transverse expansion \cite{heinz2,nu1}. The slope parameters and
the spectra have been well described by the convolution of thermal
and flow components within the framework of hydrodynamical models
\cite{models}. Thus, on the basis of these analyses one concludes
that, at the SPS, chemical and thermal equilibrium was indeed
achieved at some stage of the collision. The chemical freezeout
temperature $T_f\simeq 168\pm 2.4$ MeV found from a thermal
analysis \cite{pbm2} of experimental data in Pb-Pb collisions at
SPS  is remarkably consistent within error with the critical
temperature $T_c\simeq 170\pm 8$ MeV obtained from Lattice
Monte-Carlo simulations of QCD at vanishing baryon density
\cite{karsch1,karsch}. Thus, the observed hadrons  seem to be originating
from a deconfined medium and the chemical composition of the
system is most likely established during hadronization
\cite{stachel1,stoc}. The observed coincidence of chemical and
critical conditions in the QCD medium, if indeed valid, should
also be seen in heavy ion collisions at higher collision energies,
in particular at RHIC.

In this paper, we present  a thermal  analysis of particle
production  measured in Au-Au reactions at $\sqrt s$=130 GeV which
were reported in
\cite{starprl,harris,caines,zxu,huang,zajc,ohnishi,george,bearden,bprl}. 
We will focus on momentum integrated particle  yields and their
ratios, thereby studying   the soft component of particle spectra
which gives the major contribution to the overall multiplicity. We
derive the chemical freezeout  conditions at RHIC and discuss
their uncertainties. We also propose an experimental observable
which could further constrain the thermal parameters. Finally, we
provide
 model predictions for particle production in Au-Au reactions at
$\sqrt s=130 ~{\rm and }~200$ GeV.


\section{Results}
In the present analysis of RHIC data\footnote{ A summary of  data
of different collaborations is presented in \cite{nu}}, we adopt
the statistical model which has been used previously in the
analysis of particle production at  SPS and AGS energies
\cite{pbm2}. This model was formulated in the grand canonical
ensemble with the  baryon number, strangeness and electric charge
conservation. The partition function contains the contributions
from all mesons and baryons with masses up to 1.5 and 2.0 GeV,
respectively. To account for repulsive interactions between
hadrons at small distances, an eigenvolume was assigned to all
particles. The eigenvolume parameter is assumed to be the same for
all hadrons and is calculated with a hard core radius of 0.3 fm.
Calculations of relevant observables such as  particle densities
are performed thermodynamically in a self-consistent manner. Thus
all these quantities are derived directly from the partition
function by taking appropriate derivatives.

Within this framework particle multiplicity ratios depend only on
two independent parameters: the temperature $T$ and the baryon
chemical potential $\mu_B$. All other fugacity parameters are
determined by the initial conditions requiring   strangeness
neutrality  and electric charge conservation. The contribution of
resonance decays  to the final particle yields of lighter mesons
and baryons is also included in the analysis.

The above model is now applied to Au-Au collisions at $\sqrt s=$130 GeV at
RHIC. We use the results of the STAR \cite{starprl,harris,caines,zxu,huang},
PHENIX \cite{zajc,ohnishi}, PHOBOS \cite{george}, and BRAHMS
\cite{bearden,bprl} collaborations for different particle multiplicity ratios
listed in Table.1. All these ratios were obtained from measured particle
yields integrated  over transverse momentum. The pseudo-rapidity coverage was,
however, limited to the central region, $-0.5<\eta <0.5$.

We adjust the free parameters $T$ and $\mu_B$ to get the best description of
the data. The calculations are performed under the condition of  strangeness
neutrality within the 1 unit of rapidity covered by the data. Furthermore, we
assume that 50 \% of all weakly decaying baryons are reconstructed in the
experiments. For the criterium of the best fit we use the value of  $\chi^2$.
Fig.1 shows the  contours of constant $\chi^2$ in the $T-\mu_B$ plane. The
best fit corresponding to the minimum of $\chi^2\simeq 5.7$ for 7 effective
degrees of freedom gives $T\simeq 174\pm 7$ MeV and $\mu_B\simeq 46\pm 5$ MeV
as the chemical freezeout parameters at this RHIC energy. The corresponding
strange chemical potential is $\mu_s = 13.6 $ MeV. Repeating the calculations
assuming that as an upper limit 100 \% of all weakly decaying baryons are
reconstructed leads to $T\simeq 169\pm 7$ MeV and $\mu_B\simeq 48\pm 4$ MeV.
Since experiments presently do not correct for feed-down from weakly decaying
baryons, see, e.g., the discussion in \cite{caines}, the reconstruction
efficiency will be large, but the sensitivity of the calculations to the
actual value is small. In the absence of more detailed information we use  50
\% as reconstruction efficiency.

The energy per
particle of 1.1 GeV obtained with these parameters  is consistent
with the unified freezeout conditions proposed in \cite{prl}. We
note that these parameter values are close to those used for a
prediction of particle yields at full RHIC energy \cite{js} of
$T=168$ MeV and $\mu_B=10$ MeV..

In Fig. 2 we show the comparison  of the thermal model results
with experimental data. One sees in Fig. 2 and also in Table.1
that the overall agreement is very good. Most of the data are
reproduced by the model within the experimental errors. The
largest deviations are seen in the ratios of ${\bar K^{*0}}/h^-$
and $K^{*0}/h^-$ but still they are on the level of one standard
deviation. In Table.1 the predictions of the model for different
particle ratios which are still not known experimentally are also
presented \footnote{These calculations are made assuming a 50$\%$
reconstruction efficiency for particles resulting from weak baryon
decay. This may need to be adapted to specific experimental
conditions when  comparing ratios involving such particles to
data.}. As is visible from Fig.1, the minimum in the
$\chi^2$ surface is  narrower in $\mu_B$ direction while  more
freedom for variation is still possible in $T$ direction (see also
results presented in \cite{becattini,nu,redlich,rafelski}). To
further constrain the value of the freezeout temperature
additional results for particle ratios are required. Of particular
relevance would be the ratios of multistrange baryons to mesons
which are strongly $T$ dependent while being only weakly
influenced by $\mu_B$. In Fig.3 we show the calculated temperature
dependence of the $\Omega^- /\pi^-$, $\Xi^+/\pi^+$ and  $\Xi^-
/K^-$ ratios as possible observables which would constrain further
the value of $T$.

Considering the best values of thermal parameters derived from the present
analysis  together with  previous systematics on freezeout conditions in heavy
ion collisions  in the energy range from SIS up to SPS \cite{prl,CLE99}, one
can make  predictions for particle production in Au-Au reactions at $\sqrt
s=200$ GeV. Indeed in \cite{new} it was shown that the energy dependence of
the baryon chemical potential can be parameterized phenomenologically as
$\mu_b\sim 1.3{\rm GeV}\dot (1+\sqrt{s}/4.5{\rm GeV})^{-1}$. This prediction
is consistent within error  with our present result  for RHIC energy. Using
this parameterization   one finds a decrease of baryon chemical potential from
$\mu_B\simeq 46\pm 5$ MeV at $\sqrt s=130$ GeV to $\mu_B\simeq 29\pm 8$ MeV at
$\sqrt s=200$ GeV.
The unified freezeout condition of fixed energy/particle $\simeq
1.1$ GeV provides a temperature increase to 177$\pm 7$ MeV in
Au-Au reactions at $\sqrt s=200$. The freezeout temperature  is
obviously bounded from above by the deconfinement phase transition
temperature. The $\chi^2$ fit provides only statistical indication
of the most probable value of thermal parameters.
 In Table I we show the results
of the statistical model for different particle ratios at  the top
RHIC energy. It is clear that the small baryon and corresponding
strange chemical potentials imply that  particle to  anti-particle
ratios are close to unity. Other ratios are also at  values  near
the baryon-free limit.

Finally, we have to stress that  at energies where baryon stopping
is dominant, such as at the SPS, the comparison of the statistical
model with experiment should be done with 4$\pi$ particle yields
\cite{pbm2,xu}. This procedure strongly reduces the possible
influence of dynamical effects on particle yields and guarantees
that the conservation laws of quantum numbers are fulfilled. In
the restricted acceptance near midrapidity one needs to account
for  additional uncertainties in the derivation of thermal
conditions from the experimental data. It is not excluded that
thermal parameters at midrapidity could deviate from their values
in full phase space as  already seen at the SPS energy
\cite{becattini}. On the other hand, if the energy is sufficiently
high such that data exhibit boost-invariant rapidity plateaus,
analysis near mid-rapidity should also be little influenced by
dynamical effects such as, e.g., hydrodynamical flow \cite{cr}.
The current energy seems already to be close to this regime.

\section{Summary and conclusions}

In conclusion, we have  shown that the statistical model in complete
equilibrium gives results  consistent with  the experimental data for
particle  production in Au-Au collisions at $\sqrt s$=130GeV. At this energy
the chemical  freezeout appears at  $T= 174\pm 7$ MeV and $\mu_B = 46\pm 5$
MeV. The resulting temperature is only slightly higher than that previously
found at the SPS where for Pb-Pb collisions $T=168\pm 5$MeV.  This relatively
moderate increase of temperature could be expected since in the limit of
vanishing baryon density the temperature should not exceed the critical value
required for deconfinement. The substantial decrease of the baryon chemical
potential from $\mu_B\simeq 270$ MeV at SPS to $\mu_B\simeq 45 $ MeV  at RHIC
found in these calculations shows that at midrapidity we are dealing with a low
net baryon density  medium.

We have also discussed   possible uncertainties  of the results
and proposed observables which could provide  better constraints
on thermal parameters. Finally,  extrapolating the values   of
thermal parameters  to a higher energy, we have made a prediction
for particle production in Au-Au reactions at $\sqrt s =200$ GeV.

\section{Acknowledgements}
On of us (K.R.)  acknowledges a stimulating discussions with  J.
Cleymans, W. N\"orenberg   and  Nu Xu  as well as the partial
support of the Polish Committee for Scientific Research (KBN-2P03B
03018).
%
%


%
%
%
\newpage
\begin{figure}
\hspace*{1cm}
\includegraphics[width=.9\textwidth]{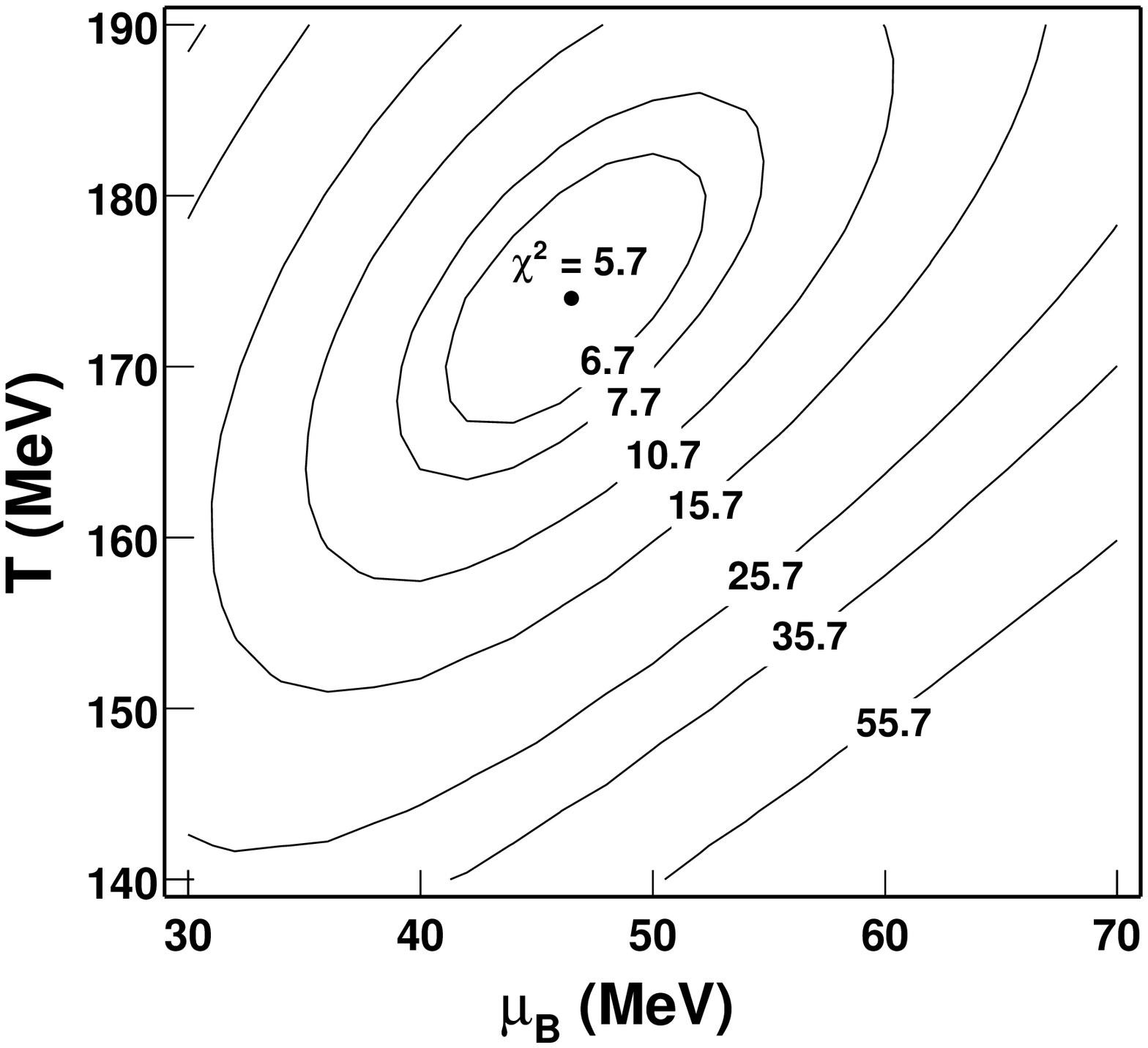}
\caption{{ Lines of constant $\chi^2$ for the comparison of experimental
particle ratios from RHIC experiments and model calculations for a wide range
of thermal parameters $(T,\mu_B)$. The dot represents the parameter set with
minimum $\chi^2$ located at $T=174$ {\rm MeV} and $\mu_B=46$ {\rm MeV}. }}
\label{mubT_e}
\end{figure}
\newpage
\begin{figure}
\hspace*{1cm}
\includegraphics[width=.9\textwidth]{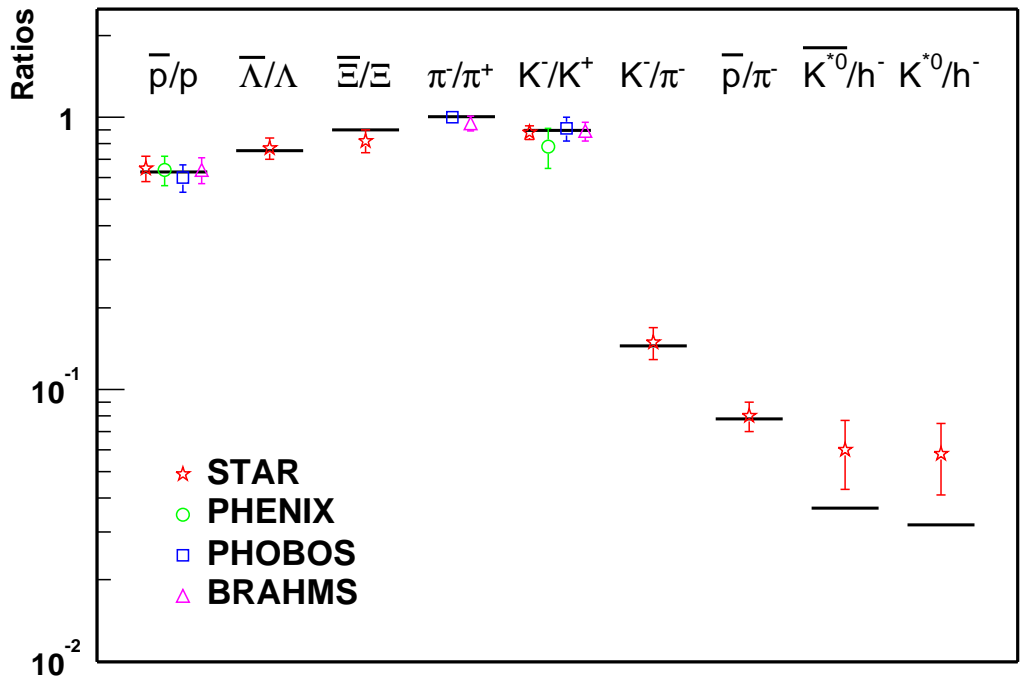}
\caption{{ Comparison between RHIC experimental particle ratios and
statistical model  calculations with $T=174$ {\rm MeV} and $\mu_B=46 $ {\rm
MeV}. Experimental data are taken from
\cite{starprl,harris,caines,zxu,huang,zajc,ohnishi,george,bearden,bprl}. }}
\label{comp}
\end{figure}
\newpage

\begin{figure}
\hspace*{1cm}
\includegraphics[width=.9\textwidth]{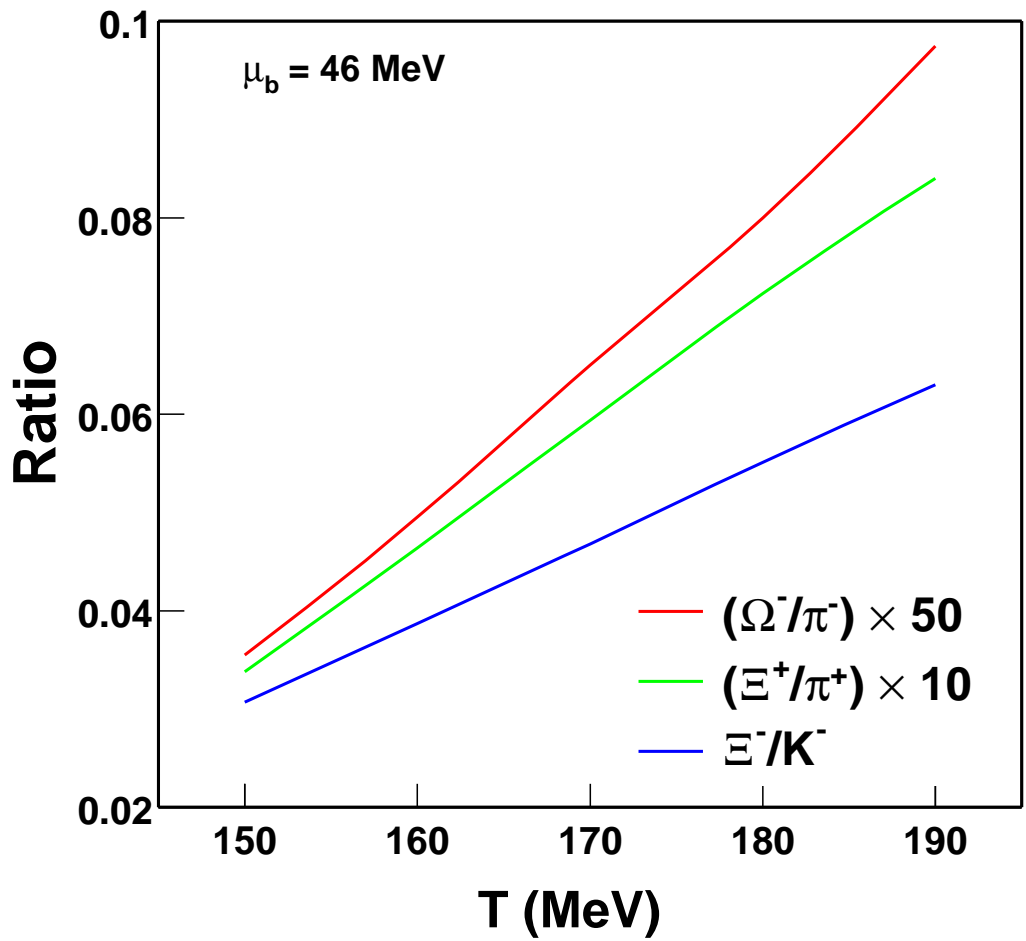}
\caption{{ Temperature dependence of different particle ratios calculated with
$\mu_B=50 $ {\rm MeV}.  }} \label{temp_dep}
\end{figure}
\newpage

\begin{table}
\begin{tabular}{l l l l l l}
Ratio                        & Model calculation  & Expt. Data   &  Expt. & Ref. \\  %
               & $\sqrt{s}=130\ (200)$ GeV & $\sqrt{s}=130$ GeV  & &  \\ \hline %

$\overline{p}/p$        & 0.629 (0.752)& 0.65 $\pm$ 0.07   & STAR & \cite{starprl} \\ %
                        &       & 0.64 $\pm$ 0.07 & PHENIX & \cite{zajc} \\ %
                        &       & 0.60 $\pm$ 0.07 & PHOBOS & \cite{george} \\ %
                        &       & 0.64 $\pm$ 0.07 & BRAHMS & \cite{bprl} \vspace{0.2cm} \\ %

$\overline{p}/\pi^{-}$  & 0.078 (0.089)& 0.08 $\pm$ 0.01   & STAR & \cite{harris} \vspace{0.2cm} \\ %

$\pi^{-}/\pi^{+}$       & 1.007 (1.004)& 1.00 $\pm$ 0.02 & PHOBOS & \cite{george} \\ %
                        &       & 0.95 $\pm$ 0.06 & BRAHMS & \cite{bearden}  \vspace{0.2cm}\\ %

$K^{-}/K^{+}$           & 0.894 (0.932)& 0.88 $\pm$ 0.05   & STAR & \cite{caines} \\ %
                        &       & 0.78 $\pm$ 0.13 & PHENIX & \cite{ohnishi}     \\ %
                        &       & 0.91 $\pm$ 0.09 & PHOBOS & \cite{george} \\ %
                        &       & 0.89 $\pm$ 0.07 & BRAHMS & \cite{bearden}  \vspace{0.2cm} \\ %

$K^{-}/\pi^{-}$         & 0.145 (0.147)& 0.149 $\pm$ 0.02  & STAR & \cite{caines} \vspace{0.2cm} \\ %

$K^{\star0}/h^{-}$          & 0.037 (0.036)& 0.06 $\pm$ 0.017  & STAR & \cite{zxu} \vspace{0.2cm} \\ %

$\overline{K^{\star0}}/h^{-}$ & 0.032 (0.033)& 0.058 $\pm$ 0.017 & STAR & \cite{zxu} \vspace{0.2cm} \\ %

$\overline{\Lambda}/\Lambda$  & 0.753 (0.842)& 0.77 $\pm$ 0.07   & STAR & \cite{zxu}  \vspace{0.2cm} \\ %

$\overline{\Xi}/\Xi$          & 0.894 (0.942)& 0.82 $\pm$ 0.08   & STAR & \cite{huang} \vspace{0.2cm} \\ %

$\Lambda/h^{-}$               & 0.040 (0.039) &       &  &  \\ %
$\Lambda/K^{\star0}$           & 1.086 (1.091) &       &  &  \\ %
$\Xi^{-}/\Lambda$          & 0.123 (0.127) &       &  &  \\ %
$\Xi^{-}/K^{-}$         & 49.8$\cdot 10^{-3}$ (50.2$\cdot 10^{-3}$) &       &  &  \\ %
$\Xi^{+}/\overline\Lambda$ & 0.145 (0.142) &       &  &  \\ %
$\Xi^{+}/\pi^{+}$       & 6.51$\cdot 10^{-3}$ (7.01$\cdot 10^{-3}$) &       &  &  \\ %
$\Omega/\Xi$               & 0.196 (0.197) &       &  &  \\ %
$\Omega^{+}/\Omega^{-}$    & 0.898 (0.941) &       &  &  \\ %
$\Omega^{-}/\pi^{-}$    & 1.47$\cdot 10^{-3}$ (1.50$\cdot 10^{-3}$) &       &  &  \\ \hline %

\end{tabular}
\caption{Comparison of experimental particle ratios and thermal model
calculations for $T=174$ MeV, $\mu_B=46$ MeV at $\sqrt{s}=130$ GeV. Also shown
in parentheses are model predictions for particle ratios at $\sqrt{s}= 200$
GeV with $T=177$ MeV, $\mu_B=29$ MeV.}
\end{table}

\end{document}